\documentclass[12pt]{JHEP}

\font\blackboard=msbm10 at 12pt
\font\blackboards=msbm7
\font\blackboardss=msbm5
\newfam\black
\textfont\black=\blackboard
\scriptfont\black=\blackboards
\scriptscriptfont\black=\blackboardss

\newcommand{\NP}{{\em Nucl.\ Phys.\ }}

\newcommand{\PL}{{\em Phys.\ Lett.\ }}

\newcommand{\PRL}{{\em Phys.\ Rev.\ Lett.\ }}

\newcommand{\gone}[1]{}

\title{Mass generation from tachyon condensation for vector fields
on D-branes}
\author{Washington Taylor\\
{Center for Theoretical Physics} \\
{MIT, Bldg.  6-306} \\
{Cambridge, MA 02139, U.S.A.} \\
{\tt wati@mit.edu}}

\abstract{The level truncation approach to string field theory is used
to study the zero-momentum action for vector excitations on a
bosonic D-brane which has been annihilated by tachyon condensation.
It is shown that in the true vacuum the translation zero modes
associated with transverse scalars on the D-brane are lifted by
spontaneous generation of mass terms.  Similarly, the U(1) gauge field
on the brane develops a nonzero mass term.}

\preprint{MIT-CTP-3010, hep-th/0008033}
\begin{document}

\section{Introduction}

It was recently proposed by Sen \cite{Sen-universality} that the decay
of unstable D-brane configurations in bosonic and type II string
theory can be systematically described using string field theory.  As
a quantitative test of this prediction, Sen suggested that the energy
gap between an unstable brane configuration and the vacuum should be
precisely computable using tree level string field theory.  Subsequent
work has verified this conjecture in a variety of contexts.
Calculations using level-truncated string field theory to analyze the
decay of the bosonic D-brane have given an energy gap which agrees
with the brane energy to 99.91\%
\cite{ks-open,Sen-Zwiebach,Moeller-Taylor}; a suggestion for
streamlining these calculations was made in \cite{Rastelli-Zwiebach}.
Similar calculations for unstable D-branes in superstring field theory
have shown agreement to a level of approximately 90\%
\cite{Berkovits-tachyon,bsz,dsr,Iqbal-Naqvi}.  D0/D4 brane
configurations in superstring field theory with a background B field
were considered in \cite{David-04}.  Lower-dimensional D-branes have
been constructed as solitonic tachyon configurations on the bosonic
D25-brane \cite{Harvey-Kraus,djmt,msz}.  It has been shown that the
string field theory analysis simplifies significantly in the limit of
a large B field \cite{dmr,hklm,Witten-tachyons,Sochichiu} and for
p-adic strings \cite{Ghoshal-Sen}.  Fundamental string excitations
have been constructed in the large B limit \cite{hklm}.  A discussion
of background independence in the context of tachyon condensation
recently appeared in \cite{Seiberg-note-tachyon}.  It has also been
suggested that renormalization group analysis of conformal field
theory may provide a simpler approach to some tachyon condensation
calculations of this type \cite{hkm}.

This set of results provides some of the first concrete evidence that
string field theory can be used as a practical calculational tool to
study nonperturbative aspects of string physics, particularly in the
case of processes involving creation and annihilation of D-branes,
which are essentially nonperturbative phenomena from the string theory
point of view.  Recent arguments
\cite{Witten-K,Horava-K,dmw-k,Witten-overview} that R-R charge is best
described through K-theory, which arises naturally in the context of
the annihilation through tachyon condensation of an infinite number of
branes and anti-branes, indicate that further development of the
technology of string field theory may provide very useful clues to the
fundamental structures underlying string theory and M-theory.

When a single unstable D-brane (or a brane and antibrane of type II
string theory) annihilate by tachyon condensation, there is a puzzle
involving the fate of the U(1) gauge field living on the brane
\cite{Srednicki-IIB,Witten-K}.  In particular, from the point of view
of the field theory living on the brane there does not seem to be any
obvious mechanism for removing this field from the theory after the
brane annihilation process has completed.  This puzzle has been
discussed by numerous authors; one particularly interesting suggestion
was made in \cite{Yi-membranes,bhy}, where it was argued that the gauge
theory becomes confining in the new vacuum.

In this note we address this puzzle by calculating the term quadratic
in the U(1) field $A_\mu$ on a bosonic D-brane after decay to the
stable vacuum using the level truncation approach to string field
theory.  Surprisingly, we find that this quadratic term has a nonzero
coefficient, indicating that stringy effects give a mass to the U(1)
vector field in the stable vacuum.  The possible appearance of such
mass terms was discussed previously in \cite{KS-mass}.  We do not have
control at this point over the kinetic term for the vector field, so
we phrase our discussion in terms of the lifting of the zero modes on
the D-brane associated with translation invariance in a perpendicular
direction for a brane of dimension $p < 25$.  These zero modes are
T-dual to the zero modes associated with constant gauge fields on the
D25-brane.  The appearance of a quadratic term for these fields
indicates that as the D-brane is annihilated, at some point the
translation/gauge zero modes of the configuration are lifted by
stringy effects.  This gives a natural resolution of the U(1) puzzle,
in which stringy effects spontaneously give a mass to the gauge field
in the true vacuum.

In Section \ref{sec:review} we review the string field theory analysis
of tachyon condensation on a bosonic D-brane and the structure of the
effective potential for the tachyon field.  In Section
\ref{sec:lifting} we give numerical evidence that the U(1) field on
the D-brane has a nonzero mass in the stable vacuum.  Section
\ref{sec:discussion} contains a discussion of this result. 

\section{Tachyon condensation on bosonic D-branes}
\label{sec:review}

In this section we briefly review the string field theory description
of the annihilation of a bosonic D-brane, following
\cite{ks-open,Sen-universality,Sen-Zwiebach,Moeller-Taylor}.  We
follow the notation and conventions of \cite{ks-open,Moeller-Taylor}.

Witten's cubic open string field theory action is \cite{Witten-SFT}
\begin{equation}
S = \frac{1}{2 \alpha'}  \int \Phi \star Q \Phi + \frac{g}{3!}  \int \Phi \star
\Phi \star \Phi,
\label{eq:SFT-action}
\end{equation}
where $Q$ is the BRST operator and $\star$ is the string field theory star
product.
In Feynman-Siegel gauge, the string field $\Phi$ has the expansion
\begin{equation}
\Phi = \left( \phi + A_\mu \alpha^\mu_{-1}
 + \frac{1}{\sqrt{2}} B_{\mu \nu} \alpha^\mu_{-1} \alpha^\nu_{-1} +
 \beta b_{-1} c_{-1}  + \cdots \right)| 0 \rangle
\label{eq:expansion}
\end{equation}
where $| 0 \rangle = c_1 | \Omega \rangle$ is the state in the string
Hilbert space associated with the tachyon field.  
Explicit descriptions of the vertex operator for the cubic terms in
the action (\ref{eq:SFT-action})
were
given in the oscillator basis in \cite{Gross-Jevicki-12,cst,Samuel}.

For the purposes of determining the Lorentz-invariant stable vacuum,
it is sufficient to restrict attention to the scalar fields $\psi^i$
in the string field, in terms of which we write
\begin{equation}
\Phi_{\rm scalar} = \sum_{i = 1}^{\infty}  \psi^i | s_i \rangle
\label{eq:expand-scalars}
\end{equation}
where $| s_i \rangle$ are all the scalar states in ${\cal H}$.  
(We write the tachyon field as $\psi^1 = \phi$.)
The zero momentum action for the scalar string fields can be written as
\begin{equation}
V = \sum_{i, j} d_{ij} \psi^i \psi^j +
g \kappa \sum_{i, j, k} t_{ijk} \psi^i \psi^j \psi^k
\label{eq:potential}
\end{equation}
where $g$ is the string coupling constant and
\[
\kappa = \frac{3^{7/2}}{2^7} \,.
\]
The quadratic and cubic coefficients $d_{ij}$ and $t_{ijk}$ appearing
in this potential can be explicitly computed for any finite number of
terms.  The formalism needed to perform these calculations is reviewed
in \cite{WT-SFT} using notation and conventions which we follow in
this paper.

The level truncation approach to string field theory involves
systematically truncating the string field theory action by including
fields up to some fixed level $L$ (where the tachyon is taken to have
level 0), and interactions up to a total level $I$.  Level truncation
has been shown to give a good systematic approximation to string field
theory for particular calculations such as the energy of the stable
vacuum.  In \cite{ks-open,Sen-Zwiebach}, the terms in
(\ref{eq:potential}) were calculated up to truncation level (4, 8).
The equations of motion for the fields $\psi^i$ were solved in this
approximation, and it was shown that the resulting vacuum energy
differs from the perturbative vacuum energy by $98.6\%$ of the
25-brane energy $ (2 \pi^2 g^2)^{-1}$ predicted by Sen.  All the terms
in (\ref{eq:potential}) up to truncation level (10, 20) were
determined in \cite{Moeller-Taylor}, and it was shown that in this
approximation the energy gap between the stable and perturbative vacua
agrees with Sen's prediction to $99.91\%$.  The values $\langle \psi^i
\rangle$ of the scalar fields up to level 10 were determined in 
truncations up to (10, 20) to 10 decimal places; we will use these
values in the calculations of this paper.

One way of thinking about the process of tachyon condensation is in
terms of the effective potential $V (\phi)$ which arises from
integrating out all the scalar fields $\psi^i, i > 1$ from
(\ref{eq:potential}).  In terms of this effective potential, the
stable vacuum arises at $\phi_*$ where $V' (\phi_*)= 0$, and Sen's
conjecture states that
\begin{equation}
V (0) -V (\phi_*) = \frac{1}{2 \pi^2 g^2}\, .
\end{equation}
In \cite{Moeller-Taylor}, this effective potential was studied both
from the point of view of its perturbative expansion
\begin{equation}
V (\phi)  = \sum_{n = 2}^{\infty} c_n (g \kappa)^{n-2}\phi^n
 = -\frac{1}{2}\phi^2 + g \kappa \phi^3+ \cdots
\end{equation}
and nonperturbatively using numerical techniques to solve the
equations of motion arising from (\ref{eq:potential}) in the
level-truncated theory for all fields except $\phi$.  One particularly
interesting feature of this effective potential is that at every level
of truncation up to (10, 20) a branch point appears near $\phi \approx
-0.25/g$.  The effective potential seems to have a critical point near
this value.  The physical meaning of this critical point is unclear,
although some speculations were made in
\cite{Moeller-Taylor,Sen-Zwiebach-marginal}.  The existence of this
critical point does, however, indicate a finite radius of convergence
for the effective potential $V (\phi)$.  The stable vacuum $\phi_*$
lies outside this radius of convergence, although $\phi_*$ lies in the
opposite direction on the $\phi$ axis and seems to be within the
region where the branch of the effective potential containing $\phi =
0$ is well defined.

\section{Lifting of zero modes}
\label{sec:lifting}

One of the most important outstanding questions related to D-brane
annihilation by tachyon condensation is the structure of the theory in
the stable vacuum.  Since in this vacuum the D-brane has completely
disappeared from the picture, we do not expect to see open string
modes in this vacuum.  According to Sen's conjecture the kinetic terms
for all the open string modes should vanish in the stable vacuum,
essentially decoupling all massive string modes from the theory.  How
the fate of the massless U(1) field on the brane can be understood is,
however, a puzzle which has been discussed from several points of view
\cite{Srednicki-IIB,Witten-K,Yi-membranes,bhy} but not fully resolved.

A full string field theory analysis of the fate of the massless vector
field on the brane after tachyon condensation would involve a
complicated calculation involving subtle momentum-dependent factors.
Even without considering momentum-dependence, however, there is an
interesting calculation which seems to shed light on the U(1) puzzle.
The significance of this calculation is most clear in a T-dual
setting, where we consider a bosonic D$p$-brane with $p < 25$.  Under
T-duality in $k$ compact directions, the components of the $U(1)$ field in the
compact directions become transverse scalar fields $X^i$.  For a
D$p$-brane in noncompact space, the string field theory action is the
same as for a D25-brane, with the differences that the fields all have
momentum only in $p +1$ dimensions of space-time and that the
transverse scalars $X^i$ have the physical interpretation of
describing the position of the D-brane in the transverse dimensions.

Shifting the position of a D$p$-brane in one of the transverse
directions corresponds, then, to turning on a translation zero mode of
the theory associated with a constant field $X^i$.  The physical
question we will address with the calculation in this section is
whether these translation zero modes still exist in the stable vacuum
arising from tachyon condensation on a bosonic D$p$-brane with $p <
25$.  In order to answer this question, we only need to calculate the
coefficient of the quadratic term $X^i X^i$ in the effective action
arising from integrating out all massive string fields in the stable
vacuum.  We find that indeed this term seems to have a nonzero
coefficient.  This indicates that the translation zero modes are
lifted by stringy effects in the stable vacuum.  While we perform our
calculations in a fixed gauge, this physical result is
gauge-independent.

Although the physical interpretation of this mass term is clearest in
the T-dual context of a D$p$-brane $p < 25$, we carry out this
calculation in the context of the zero-momentum action on a D25-brane.
In this picture, the quadratic term we derive represents a
spontaneously generated mass for the vector field $A_\mu$ in the
stable vacuum.  We discuss the physical meaning of this term and the
question of gauge invariance in the following section; in this section
we simply perform the string field theory calculation of the term in
question.

To make the discussion precise, we introduce some new notation.  We
are interested in the zero-momentum string field theory action for all
string fields at odd levels which transform as Lorentz vectors.  We
write this vector part of the string field as
\begin{eqnarray}
\Phi_{\rm vector}  & = & 
 \sum_{m = 1}^{\infty}  \eta^{(m)}_\mu | v_{(m)}^\mu \rangle  \nonumber\\
& = & A_{\mu} \alpha^\mu_{-1}| 0 \rangle +
\eta^{(2)}_\mu\alpha^\mu_{-1} (\alpha_{-1} \cdot \alpha_{-1})| 0
 \rangle + \label{eq:expand-vector}\\
& &\hspace{0.2in}
\eta^{(3)}_\mu\alpha^\mu_{- 3}| 0
 \rangle+\eta^{(4)}_\mu\alpha^\mu_{-1} b_{-1} c_{-1}| 0
 \rangle+ \cdots \nonumber
\end{eqnarray}
where $\eta^{(m)}_{\mu}$ are the odd-level vector fields in the theory, with
$\eta^{(1)}_{\mu} = A_\mu$.
We are interested in the zero-momentum scalar-vector-vector couplings
in the string field theory action
\begin{equation}
\kappa g\, C_{imn}\, \psi^i \left(\eta^{(m)} \cdot \eta^{(n)} \right)
\end{equation}
and the zero-momentum mass terms for the vector fields in the
perturbative vacuum
\begin{equation}
D_{mn} \left(\eta^{(m)} \cdot \eta^{(n)} \right)
\end{equation}

In terms of these couplings, the complete set of zero-momentum
quadratic terms for the vector fields in the stable vacuum can be
written as
\begin{equation}
S_{\rm v2} =  \sum_{m, n}
M_{mn}
\left(\eta^{(m)} \cdot \eta^{(n)} \right)
\end{equation}
where
\begin{equation}
M_{mn} = D_{mn} + g \kappa \sum_{i} C_{imn} \langle \psi^i \rangle 
\end{equation}
is the mass matrix for the vector fields.

If the zero modes of $A_\mu$  persist in the stable vacuum, we would
expect that in the full theory the mass matrix $M_{mn}$ would have a
vanishing eigenvalue.  We can test this using the level truncation
method by truncating at a sequence of levels $(L_v, L_s, I)$ where
$L_v$ and $L_s$ are the maximum levels of vector fields and scalar
fields considered, and $I$ is the maximum total level considered in
any interaction.  We describe here  explicitly the mass matrix in the level
truncations (1, 0, 2), (1, 2, 4) and (3, 4, 8).
\vspace{0.1in}

\noindent
{\bf Level (1, 0, 2) truncation:}

There is only a single scalar at level zero ($\phi$) and a single
vector at level one ($A_\mu$).  The complete zero-momentum string field
theory action for these fields is
\begin{equation}
S_{1, 0, 2} = -\frac{1}{2}\phi^2 + \kappa g \phi^3
+\frac{16}{9}  \kappa g \phi A_\mu A^\mu.
\end{equation}
At this level the mass matrix is
\begin{equation}
M = \left( \frac{16}{9}  \kappa g \langle \phi \rangle \right)
\approx (0.59259)
\end{equation}
when we use the level (0, 0) approximation $g \kappa\langle \phi
\rangle_{(0, 0)} = 1/3$.  Thus, in the level (1, 0, 2) truncation the
smallest eigenvalue of $M$ is $\approx 0.59$.
\vspace{0.1in}

\noindent
{\bf Level (1,  2,  4) truncation:}

There are two additional scalars appearing at level 2.  Including
these scalars the mass matrix becomes
\begin{equation}
M = \left( \kappa g \left[\frac{16}{9}  \langle \phi \rangle
- \frac{1568}{243}  \langle \psi^2 \rangle
-\frac{176}{243}  \langle \psi^3 \rangle \right]
\right)\approx (0.67289)
\end{equation}
using the level (2, 4) approximation to the scalar expectation
values.  We see that including scalars at higher level has shifted the
mass of the vector slightly, but in an upward direction.
\vspace{0.1in}

\noindent
{\bf Level (3, 4, 8) truncation:}

Using the level (4, 8) approximations for the 10 scalars at level 4 or
less we find a mass matrix for the four vectors appearing in
(\ref{eq:expand-vector}) given approximately by
\begin{equation}
M \approx \left(\begin{array}{cccc}
0.67910& -2.82440& -0.08261& 
  -0.26102\\
 -2.82440& 82.15982& 
  0.31499& 1.03418\\
 -0.08261& 
  0.31499& 4.70145& 0.02849\\ 
 -0.26102& 1.03418& 0.02849& 
  -0.95150
\end{array} \right)
\label{eq:m34}
\end{equation}
The smallest eigenvalue of this matrix (in absolute value) is $\approx
0.61206$, and the associated eigenvector is  $\approx(0.98919,
0.03598, 0.01819, -0.14101)$.
\vspace{0.1in}

We have continued this analysis of the zero-momentum mass matrix  up
to level (9, 10, 20).  The results for the smallest eigenvalue of $M$
are given in Table~\ref{t:results}.  As is apparent from the table,
there is no sign  of a vanishing eigenvalue in the limit of large
level.  Quite to the contrary, the smallest eigenvalue seems to be
converging to a number in the vicinity of $0.59$.  For each of the
eigenvalues given in  Table~\ref{t:results}, the associated
eigenvector is dominated by the component in the direction of $A_\mu$,
just as in the (3, 4, 8) example given explicitly above.

\TABLE{
\begin{tabular}{| | r | | c | c | |c | c | |}
\hline
\hline
 level & $N_v$ & $N_s$ & $\mu_{\rm min}$ & $G (\phi_*) $\\
\hline
\hline
(1, 2, 4) &1 & 3 & 0.67289&0.67289\\
\hline
(3, 4, 8) & 4 & 10 &0.61206& 0.63343\\
\hline
(5, 6, 12) &17 & 31 &0.59906& 0.62738\\
\hline
(7, 8, 16) &61 & 91 &0.59333 &0.62491\\
\hline
(9, 10, 20) &197 &252 &0.59026 & 0.62362\\
\hline
\hline
\end{tabular}
\caption{\footnotesize $N_v, N_s$ are numbers of vector and scalar
fields in level truncation $(L_v, L_s, I)$.  $\mu_{{\rm min}}$ is
smallest eigenvalue of mass matrix $M_{mn}$.  $G (\phi_*)$ is mass of
vector field in effective action for $\phi, A_\mu$.}
\label{t:results}
}

One way of physically interpreting the mass matrix for the vector
fields $\eta^{(n)}$ is in terms of the effect of this mass matrix on
the effective action for the tachyon field $\phi$ and the massless
vector $A_\mu$.  Integrating out all fields but $\phi$ and $A_\mu$
gives an effective action of the form
\begin{equation}
S_{\phi, A} = V (\phi) + G (\phi) A_\mu A^\mu +{\cal O} (A^4)
\label{eq:effective-pa}
\end{equation}
where $V (\phi)$ is the effective tachyon potential discussed in
Section 2 and $G (\phi)$ is another function of $\phi$.  The quadratic
term for the vector field in the stable vacuum has a coefficient  $G
(\phi_*)$.

We can determine $G (\phi_*)$ in terms of the matrix $M$ by solving
the equations of motion for the vector fields at leading order.  These
equations of motion state that
\begin{equation}
M_{mn} \eta^{(n)}_{\mu} = 0, \;\;\;\;\;  m > 1\,.
\end{equation}
Thus, we have
\begin{equation}
\eta^{(n)}_\mu =  (M^{-1})_{n1} \alpha_\mu
\end{equation}
for some  $\alpha_\mu$, so
\begin{equation}
\eta^{(n)}_{\mu} = \frac{(M^{-1})_{n1}}{(M^{-1})_{11}}  A_\mu,
\end{equation}
so
\begin{equation}
G (\phi_*) = \frac{1}{(M^{-1})_{11}} \,.
\end{equation}
For the examples above with $L_v = 1$, this simply gives $G (\phi_*)=
M_{11}$ as expected.  For the matrix  (\ref{eq:m34}) describing level
truncation (3, 4, 8), this gives
\begin{equation}
G (\phi_*)_{(3, 4, 8)} \approx 0.63343\,.
\end{equation}
This quantity was also calculated by Sen and Zwiebach in the course of
their analysis of marginal deformations of the bosonic D-brane away
from the perturbative vacuum \cite{Sen-Zwiebach-marginal}.  Using an
independent computational technique they arrived at precisely the same
value in this level truncated theory\footnote{I would like to thank
these authors for comparing the details of this calculation and for
extended discussions regarding its significance.}.  We have extended
this calculation of $G (\phi_*)$ up to the level (9, 10, 20)
truncation for which we have determined $M_{mn}$.  The results of this
analysis are summarized in Table~\ref{t:results}.  We see that
$G (\phi_*)$ seems to approach a nonzero number near $0.62$ in the
limit of large levels, indicating that a mass for the vector field is
spontaneously generated in the stable vacuum.  The rate at which $G
(\phi_*)$ converges to its limiting value is similar to the rate at
which $V (\phi_*)$ approaches its limiting value as higher levels are
incorporated; for example, the difference in $G (\phi_*)$ between the
level (7, 8, 16) and level (9, 10, 20) truncations is less than
$0.2\%$.  This similarity with the rate of convergence of $V (\phi_*)$
gives us confidence that the level
truncation method is giving us a good sequence of approximations to a
limiting value for $G (\phi_*)$ near 0.62.

\section{Discussion}
\label{sec:discussion}

In this note we have described a calculation which shows that the
transverse scalars $X^i$ and U(1) gauge field $A_\mu$ get mass terms
from stringy effects when a bosonic D-brane annihilates through
tachyon condensation.  In particular, we have found that in the stable
open string field theory vacuum associated with tachyon condensation
on a D$25$-brane, there is a term of the form
\begin{equation}
G (\phi_*) A_\mu A^\mu, \;\;\;\;\; G (\phi_*) \approx 0.62
\label{eq:term}
\end{equation}
in the effective potential $S(\phi, A_\mu)$.
While we performed this calculation in a particular gauge, the result
that zero modes of transverse scalars and the world-volume gauge field
are lifted in the stable vacuum is a physically observable
gauge-independent effect.

The appearance of the term (\ref{eq:term}) seems very surprising from a
field theory point of view.  In the vicinity of the perturbative
vacuum, we expect to have a massless U(1) gauge field on the brane,
and gauge invariance under this U(1) seems to guarantee the absence of
such a mass term.  How does this field theoretic picture mesh with the
structure of the string field theory action?  To understand the
connection it is helpful to consider the (non gauge-fixed) string
field theory action around the perturbative vacuum
\begin{equation}
S =  -\frac{1}{2}\phi^2 + \kappa g \phi^3
+\frac{16}{9}  \kappa g \phi A_\mu A^\mu+ \cdots
\label{eq:perturbative}
\end{equation}
It is interesting to note that the leading terms
(\ref{eq:perturbative}) appearing in the string field theory action
have also been calculated directly from the point of view of the sigma
model using a nonstandard regularization scheme \cite{kpp}.  The
action (\ref{eq:perturbative}) does not have a gauge invariance under
the transformations
\begin{eqnarray}
\delta \phi & = &  0 \label{eq:gauge}\\
\delta A_\mu & = &  \partial_\mu \Lambda \nonumber
\end{eqnarray}
Instead, (\ref{eq:perturbative}) is invariant under the BRST transformation
\begin{eqnarray}
\delta \phi & = &   \frac{32}{9} \kappa gA^\mu \partial_\mu \Lambda +
\cdots  \nonumber\\
\delta A_\mu & = &  \partial_\mu \Lambda + \cdots \label{eq:BRST}
\end{eqnarray}
After integrating out all massive fields in the theory, the effective
action $S (\phi, A_\mu)$ for the tachyon and string vector field
continues to have an invariance under a symmetry whose leading terms
are given by (\ref{eq:BRST}).  Thus, $S (\phi, A)$ is not
gauge-invariant in the sense of (\ref{eq:gauge}).  

In the neighborhood of the perturbative vacuum, it is possible to find
a field redefinition $\tilde{\phi} = f (\phi, A), \tilde{A} = g (\phi,
A)$ to a new set of fields under which the effective action $S (\tilde{\phi},
\tilde{A})$ becomes gauge invariant under (\ref{eq:gauge}).  Such a field
redefinition was recently discussed explicitly in \cite{David}.  Under
this field redefinition, the zero-momentum part of $S (\tilde{\phi},
\tilde{A})$ must become independent of $\tilde{A}$ by gauge
invariance.  From the first few terms in the effective action $S
(\phi, A)$, which coincide with (\ref{eq:perturbative}), and from
(\ref{eq:BRST}), it is clear that the first term in this field
redefinition must be
\begin{equation}
\phi \rightarrow \tilde{\phi} + \frac{16}{9} g \kappa  A_\mu A^\mu + \cdots
\label{eq:first}
\end{equation}
This redefinition cancels the term of the form $\phi A_\mu A^\mu$ in
$S (\phi, A)$.  It was furthermore shown in \cite{WT-SFT} that $S
(\phi, A)$ contains a term of the form $A_\mu A^\mu A_\nu A^\nu$ which
was determined to within a few percent using level truncation
including fields up to level 20; this term is also precisely cancelled
by (\ref{eq:first}).

Although the existence of a field redefinition of this type seems to
mean that there is a set of variables in terms of which the effective
action $S (\phi, A)$ has the standard gauge invariance
(\ref{eq:gauge}), care must be taken with respect to the range of
validity of such a field redefinition, as emphasized in \cite{kpp}.
This field redefinition is only defined in a perturbative fashion, and
therefore we expect that it will only be well-defined in a finite
sized neighborhood of the perturbative vacuum.  The apparent existence
of a nonzero mass term (\ref{eq:term}) for $A_\mu$ in the
nonperturbative stable vacuum of the open bosonic string field theory
can in fact be interpreted as evidence that the U(1) gauge theory
description of the bosonic D-brane theory has a finite range of
validity.  If this interpretation is correct, it would mean that as
the tachyon condenses and the theory moves towards the nonperturbative
stable vacuum, at a certain point the theory undergoes a phase
transition beyond which an effective gauge field theory description of
the system such as suggested in \cite{Sen-effective,Garousi,bddep} would no
longer be applicable.  This is clearly an important issue to
understand better, since recent analyses of physics in the
nonperturbative vacuum \cite{hklm,Suyama,gms} rely on extending such a
gauge-invariant field theory action all the way to the stable vacuum.
Since the results found in these papers are generally physically
correct, it may be that even if a mass term such as we have found does
exist for the U(1) field in the stable vacuum, some qualitative
aspects of the physics in this vacuum are correctly captured by a
gauge-invariant field theory.

One possible scenario which seems to be consistent with the picture
given here is that the transformation law for $A_\mu$ in the effective
theory $S (\phi, A_\mu)$ may be of the form
\begin{equation}
\delta A_\mu = f (\phi) \partial_\mu \Lambda + \cdots
\end{equation}
where $f (\phi_*)= 0$.  This seems to be the simplest way to ensure
that an action of the form (\ref{eq:effective-pa}) retains BRST
symmetry at a point where $V' (\phi_*)= 0$ and $G (\phi_*)\neq 0$.
This scenario might also help explain why the locally defined gauge
field theory comes close to capturing the physics of the true vacuum
correctly, as it is possible that the field theory picture is valid
for all $\phi < \phi_*$.

A skeptic might argue that the numerical results presented here are
rather flimsy evidence on which to base a claim that gauge theory
breaks down in the process of tachyon condensation.  It is certainly
not impossible, for example, that the numbers in
Table~\ref{t:results}, while appearing to converge to quantities in
the neighborhood of 0.6, actually are the beginning of two very long
sequences of numbers which converge slowly to 0.  This seems unlikely,
at least to the present author, since the apparent convergence of the
value of $G (\phi_*)$, in particular, seems to be proceeding in a
closely analogous fashion to the convergence of $V (\phi_*)$ as
studied in \cite{Moeller-Taylor}.  These functions arise in a very
parallel fashion in the string field theory picture, so it would seem
very surprising if the level truncation approach gives such a good
approximation to $V (\phi)$ near the stable vacuum and is nonetheless
giving misleading information for $G (\phi_*)$.  As further evidence
that the scenario outlined above is a correct description of the
physics, it is also relevant to consider the fact that the
perturbative expansion of $V (\phi)$ has a radius of convergence
noticeably smaller than $\phi_*$.  This seems to lend credence to the
possibility that the field redefinition needed to give a
gauge-invariant formulation of the D-brane theory may have a similar
radius of convergence and may not be well-defined in the vicinity of
the stable vacuum.

The results presented here seem to give a very simple solution to the
puzzle regarding the fate of the massless U(1) field, when viewed from
the point of view of string field theory.  In the picture we have
given, the U(1) field becomes massive due to stringy effects when the
tachyon condensation becomes sufficient to bring the theory near the
stable vacuum.  It is not clear whether there is a way to bring this
intrinsically stringy effect back into the domain of field theory, so
that this phenomenon might be described either in terms of a Higgs
effect or confinement as suggested in \cite{Yi-membranes,bhy}.

In this work we have neglected momentum dependence in the string field
theory action.  Including all momentum-dependent terms in the action
and constructing the full momentum-dependent quadratic string field
theory action in the vicinity of the stable vacuum would make possible
a detailed study of the pole spectrum of the theory, extending earlier
work in \cite{ks-open}, which would shed a great deal of light on the
fate of the open string excitations in the stable vacuum.  In
particular, to precisely understand the fate of the U(1) vector field
it is clearly desirable to have control over the kinetic terms of the
vector fields as well as the mass terms studied here. One likely
scenario, which was suggested by Sen \cite{Sen-universality}, is that
all the kinetic terms will vanish in the stable vacuum.  If this is
true and the quadratic matrix $M_{nm}$ for the vector fields indeed
has a lower bound on its spectrum as suggested by the results
described here, all the vector fields in the theory will become
infinitely massive and hence decouple in the stable vacuum.  This
would fit in very nicely with the expected decoupling of all open
string fields in this vacuum.  Another, more bizarre, possibility is
that the kinetic terms for the gauge field $A_\mu$ become infinite in
the effective action $S (\phi, A)$ using the standard string field
theory normalization conventions we have used here.  This would
essentially drive the mass of the gauge field back to zero even in the
presence of the mass term we have found.  This possibility would seem
to be very unlikely as it is difficult to reconcile with the
decoupling of the other massive string modes, but it is worth
mentioning since it seems to be the only possibility other than an
extraordinary conspiracy at high level that would restore the massless
U(1) field at the stable vacuum.  Work is currently in progress to
give a complete analysis of the spectrum of the level-truncated string
fields around the stable vacuum, including all momentum-dependent
factors.  When completed, such an analysis may give new insights as to
how the complete set of open string fields including the apparently
massive U(1) decouple in the stable vacuum.

Finally, it would clearly be of interest to extend the analysis in
this note to the case of the superstring.  While there seem to be
complications in using Witten's cubic superstring field theory for
this kind of calculation (see \cite{dsr2} and references therein), the
approach of Berkovits \cite{Berkovits-general} has led to successful
calculations of the energy of unstable D-branes in type II string
theory to an accuracy of about 90\%
\cite{Berkovits-tachyon,bsz,dsr,Iqbal-Naqvi}.  It would be very
interesting to see whether the world-volume U(1) field on such a brane
has a mass analogous to that computed in this note in the
supersymmetric type II vacuum.  On the one hand, it is natural to
expect that the physics of this U(1) field should not be particularly
different in the supersymmetric case from the bosonic case.  On the
other hand, however, arguments for a simple form of the action
describing the tachyon and U(1) field were given in
\cite{Sen-effective,Garousi,bddep}.  Some of these arguments relied on
supersymmetry.  It may be that supersymmetry helps to protect the U(1)
gauge field from the spontaneously generated mass we have found here
in the bosonic case.  More work in this direction is clearly needed to
clarify this issue.

\section*{Acknowledgements}

I would like to particularly thank Ashoke Sen and Barton Zwiebach for
extensive discussions on the issues discussed in this paper and for
comparing detailed aspects of their own calculations with those I have
described here.  I would also like to thank Justin David, Kentaro
Hori, Albion Lawrence and Leonardo Rastelli for helpful discussions.
Thanks also to the Aspen Center for Physics, where this work was
completed, and to the participants in the Aspen Summer '00 M-theory
and Duality Workshop for many stimulating discussions relevant to the
subject matter of this note.  This work was supported in part by the
A.\ P.\ Sloan Foundation and in part by the DOE through contract
\#DE-FC02-94ER40818.

\normalsize
\newpage

\bibliographystyle{plain}

\begin{thebibliography}{10}

\bibitem{Sen-universality}
A.\ Sen, ``Universality of the tachyon potential,'' {\it JHEP} {\bf 9912}
  (1999) 027, {\tt hep-th/9911116}.

\bibitem{ks-open}
V.\ A.\ Kostelecky and S.\ Samuel, ``On a nonperturbative vacuum for the open
  bosonic string,'' \NP {\bf B336} (1990) 263-296.

\bibitem{Sen-Zwiebach}
A.\ Sen, B.\ Zwiebach ``Tachyon condensation in string field theory,'' {\it
  JHEP} {\bf 0003} (2000) 002, {\tt hep-th/9912249}.

\bibitem{Moeller-Taylor}
N.\ Moeller and W.\ Taylor, ``Level truncation and the tachyon in open bosonic
  string field theory,'' {\tt hep-th/0002237}.

\bibitem{Rastelli-Zwiebach}
L.\ Rastelli and B.\ Zwiebach ``Tachyon potentials, star products and
  universality,'' {\tt hep-th/0006240}.

\bibitem{Berkovits-tachyon}
N.\ Berkovits, ``The tachyon potential in open Neveu-Schwarz string field
  theory,'' {\it JHEP} {\bf 0004} (2000) 022, {\tt hep-th/0001084}.

\bibitem{bsz}
N.\ Berkovits, A.\ Sen and B.\ Zwiebach, ``Tachyon condensation in superstring
  field theory,'' {\tt hep-th/0002211}.

\bibitem{dsr}
P.\-J.\ De Smet and J.\ Raeymaekers, ``Level four approximation to the tachyon
  potential in superstring field theory,'' {\em JHEP} {\bf 0005} (2000) 051,
  {\tt hep-th/0003220}.

\bibitem{Iqbal-Naqvi}
A.\ Iqbal and A.\ Naqvi, ``Tachyon condensation on a non-BPS D-brane,'' {\tt
  hep-th/0004015}.

\bibitem{David-04}
J.\ R.\ David, ``Tachyon condensation in the D0/D4 system,'' {\tt
  hep-th/0007235}.

\bibitem{Harvey-Kraus}
J.\ A.\ Harvey and P.\ Kraus ``D-branes as unstable lumps in bosonic open
  string field theory,'' {\it JHEP} {\bf 0004} (2000) 012, {\tt
  hep-th/0002117}.

\bibitem{djmt}
R.\ de Mello Koch, A.\ Jevicki, M.\ Mihailescu and R.\ Tatar, ``Lumps and
  p-branes in open string field theory,'' \PL {\bf B482} (2000) 249, {\tt
  hep-th/0003031}.

\bibitem{msz}
N.\ Moeller, A.\ Sen and B.\ Zwiebach ``D-branes as tachyon lumps in string
  field theory,'' {\tt hep-th/0005036}.

\bibitem{dmr}
K.\ Dasgupta, S.\ Mukhi and G.\ Rajesh, ``Noncommutative tachyons,'' {\it JHEP}
  {\bf 0006} (2000) 022, {\tt hep-th/0005006}.

\bibitem{hklm}
J.\ A.\ Harvey, P.\ Kraus, F.\ Larsen and E.\ J.\ Martinec ``D-branes and
  strings as noncommutative solitons,'' {\em JHEP} {\bf 0007} (2000) 042, {\tt
  hep-th/0005031}.

\bibitem{Witten-tachyons}
E.\ Witten, ``Noncommutative tachyons and string field theory,'' {\tt
  hep-th/0006071}.

\bibitem{Sochichiu}
C.\ Sochichiu, ``Noncommutative tachyonic solitons. Interaction with gauge
  field,'' {\tt hep-th/0007217}.

\bibitem{Ghoshal-Sen}
D.\ Ghoshal and A.\ Sen, ``Tachyon condensation and brane descent relations in
  p-adic string theory,'' {\tt hep-th/0003278}.

\bibitem{Seiberg-note-tachyon}
N.\ Seiberg, ``A note on background independence in noncommutative gauge
  theories, matrix model, and tachyon condensation,'' {\tt hep-th/0008013}.

\bibitem{hkm}
J.\ A.\ Harvey, D.\ Kutasov and E.\ J.\ Martinec, ``On the relevance of
  tachyons,'' {\tt hep-th/0003101}.

\bibitem{Witten-K}
E.\ Witten, ``D-branes and K-theory,'' {\it JHEP} {\bf 9812}:019 (1998), {\tt
  hep-th/9810188}.

\bibitem{Horava-K}
P.\ Horava, ``Type IIA D-branes, K-theory, and matrix theory,'' {\em Adv.
  Theor. Math. Phys.} {\bf 2}, 1373 (1999), {\tt hep-th/9812135}.

\bibitem{dmw-k}
D.\-E.\ Diaconescu, G.\ Moore and E.\ Witten, ``E (8) gauge theory, and a
  derivation of K theory from M theory,'' {\tt hep-th/0005090}; ``A derivation
  of K theory from M theory,'' {\tt hep-th/0005091}; .

\bibitem{Witten-overview}
E.\ Witten, ``Overview of K-theory applied to strings,'' {\tt hep-th/0007175}.

\bibitem{Srednicki-IIB}
M.\ Srednicki, ``IIB or not IIB,'' {\it JHEP} {\bf 08} (1998) 005, {\tt
  hep-th/9807138}.

\bibitem{Yi-membranes}
P.\ Yi, ``Membranes from five-branes and fundamental strings from
  D$p$-branes,'' \NP {\bf B550} (1999) 214, {\tt hep-th/9901159}.

\bibitem{bhy}
O.\ Bergman, K.\ Hori and P.\ Yi, ``Confinement on the brane,'' \NP {\bf B580}
  (2000) 289, {\tt hep-th/0002223}.

\bibitem{KS-mass}
V.\ A.\ Kostelecky and S.\ Samuel, ``Photon and graviton masses in string
  theories,'' \PRL {\bf 66} (1991) 1811.

\bibitem{Witten-SFT}
E.\ Witten, ``Non-commutative geometry and string field theory,'' \NP {\bf
  B268} (1986) 253.

\bibitem{Gross-Jevicki-12}
D.\ J.\ Gross and A.\ Jevicki, ``Operator formulation of interacting string
  field theory (I), (II),'' \NP {\bf B283} (1987) 1; \NP {\bf B287} (1987) 225.

\bibitem{cst}
E.\ Cremmer, A.\ Schwimmer and C.\ Thorn, ``The vertex function in Witten's
  formulation of string field theory'' \PL {\bf B179} (1986) 57.

\bibitem{Samuel}
S.\ Samuel, ``The physical and ghost vertices in Witten's string field
  theory,'' \PL {\bf B181} (1986) 255.

\bibitem{WT-SFT}
W.\ Taylor, ``D-brane effective field theory from string field theory,'' {\tt
  hep-th/0001201}.

\bibitem{Sen-Zwiebach-marginal}
A.\ Sen, B.\ Zwiebach ``Large marginal deformations in string field theory,''
  {\tt hep-th/0007153}.

\bibitem{kpp}
V.\ A.\ Kostelecky, M.\ J.\ Perry and R.\ Potting, ``Off-shell structure of the
  string sigma model,'' \PRL {\bf 84} (2000) 4541, {\tt hep-th/9912243}.

\bibitem{David}
J.\ R.\ David, ``U(1) gauge invariance from open string field theory,'' {\tt
  hep-th/0005085}.

\bibitem{Sen-effective}
A.\ Sen, ``Supersymmetric world-volume action for non-BPS D-branes,'' {\it
  JHEP} {\bf 9910} (1999) 008, {\tt hep-th/9909062}.

\bibitem{Garousi}
M.\ R.\ Garousi, ``Tachyon couplings on nonBPSD-branes and Dirac-Born-Infeld
  action,'' {\tt hep-th/0003122}.

\bibitem{bddep}
E.\ Bergshoeff, M.\ de Roo, T.\ C.\ de Wit, E.\ Eyras and S.\ Panda,
  ``T-duality and actions for non-eps D-branes,'' {\tt hep-th/0003221}.

\bibitem{Suyama}
T.\ Suyama, ``Description of intersecting branes via tachyon condensation,''
  {\tt hep-th/0006052}.

\bibitem{gms}
R.\ Gopakumar, S.\ Minwalla and A.\ Strominger, ``Symmetry restoration and
  tachyon condensation in open string theory,'' {\tt hep-th/0007226}.

\bibitem{dsr2}
P.\-J.\ De Smet and J.\ Raeymaekers, ``The tachyon potential in Witten's
  superstring field theory,'' {\tt hep-th/0004112}.

\bibitem{Berkovits-general}
N.\ Berkovits ``Super-Poincare invariant superstring field theory,'' \NP {\bf
  B450} (1995) 90, {\tt hep-th/9503099}.

\end{thebibliography}

\end{document}